# A BIBLIOMETRIC ANALYSIS OF TRUST IN CONVERSATIONAL AGENTS OVER THE PAST FIFTEEN YEARS

Meltem Aksoy[1]*, Annika Bush[1]

[1] Research Center Trustworthy Data Science and Security of the University Alliance Ruhr, Faculty of Informatics, Technical University Dortmund, Germany

**Abstract**
Conversational agents (CA) have become increasingly prevalent in various domains, driving significant interest in understanding the dynamics of trust in CA. This study addresses the need for a comprehensive analysis of research trends in this field, especially given the rapid advancements and growing use of CA technologies like ChatGPT. Through bibliometric analysis, we aim to identify key keywords, disciplines, research clusters, and international collaborations related to CA and trust. We analyzed 955 studies published between 2009 and 2024, all sourced from the Scopus database. Additionally, we conducted a text clustering analysis to identify the main themes in the publications and understand their distribution. Our findings highlight the increasing interest in CA, particularly with the introduction of ChatGPT. The USA leads in research output, followed by Germany, China, and the UK. Furthermore, there is a notable rise in interdisciplinary research, especially in the fields of human-computer interaction and artificial intelligence.

## 1. INTRODUCTION

Over the past two decades, the widespread adoption of conversational agents (CA) in various domains, from customer service bots to personal assistants and health counselors, has generated considerable interest in understanding the dynamics of human-agent interaction. A CA is an artificial intelligence (AI) system designed to interact with humans in a natural, conversational manner. These agents can vary from simple scripted chatbots that provide predefined responses to sophisticated AI-driven assistants capable of learning and adapting their responses over time. Their primary aim is to mimic human-like interactions, providing users with information, assistance, or companionship through a conversational interface. These systems permit users to perform a range of tasks via voice or written commands on a telephone or computer. This is enabled by a variety of technologies, including natural language generation, natural language understanding, machine learning (ML), speech recognition, text-to-speech translation, and dialog management.

The first known example of CA, ELIZA, was designed to mimic conversations with a therapist. Developed in 1966, this system only had simple word-matching capabilities due to the limited technology of the time. However, this concept has evolved rapidly, especially in recent years, thanks to significant advances in ML and natural language processing (NLP). These advances have made CA much more capable and sophisticated. Leading CA such as IBM Watson, Siri, Alexa, and Google Assistant demonstrate the rapid evolution and advanced potential of these systems. Although basic forms already existed since the 1960s, there has been a huge development of CA in recent years (Deng and Lin, 2022; Radziwill and Benton, 2017). The release of ChatGPT by OpenAI in 2022 in particular has brought CA from a niche topic into the public eye and ever since the global and public interest in AI and CA has been higher than ever. According to the main functionality, CA is divided into two categories: task-oriented agents and social agents. Task-oriented agents are designed to perform specific tasks, such as booking a hotel or ordering food, and often operate within a limited domain (Wahde and Virgolin, 2022). Social agents, on the other hand, are designed to engage in more open-ended conversations, often for companionship or entertainment. These distinctions are not always clear-cut, as the capabilities of CA continue to evolve (Bickmore and Cassell, 2005; Gupta et al., 2019; Sadek et al., 2023).

We decided to investigate CA due to their impact on human-technology interaction, especially human-AI interaction, and, therefore, people's understanding and acceptance of new AI-based technologies. CA, including chatbots and virtual assistants, are advancing rapidly due to improvements in AI, NLP, and ML, and have a significant impact on e.g. everyday life, research, industry, and businesses. A bibliometric analysis can help track the evolution of these technologies, identify key innovations, and highlight emerging trends and future directions.

We are especially interested in the human-centered research of CA and, therefore, decided on a key topic in human-technology interaction research: trust. Trust has emerged as a central theme in the study of CA since it is a key element in human-computer interaction (Wischnewski et al., 2023). Trust influences not only the initial acceptance of these technologies but also their ongoing use (Cai et al., 2022; Cheng et al., 2022; Gupta et al., 2022; Rheu et al., 2021). The literature abounds with studies exploring various aspects of 'Trust in CA', e.g. examining how trust is formed, the factors that influence it, and its impact on user behavior (e.g. Andrés-Sánchez and Gené-Albesa, 2024; Andries and Robertson, 2023; Chen et al., 2023; Hofeditz et al., 2023). These studies provide valuable insights into designing more effective and trustworthy CA. Also, there are systematic literature review studies summarizing research on trust and CA (Ling et al., 2021; Loveys et al., 2020; Rheu et al., 2021). However, there is a notable gap in comprehensive bibliometric analyses that identify trends and core topics within this field. To fill to this literature gap, our study provides a comprehensive overview of research done on 'Trust in CA' over the last 15 years using bibliometrics analysis.

Bibliometric analysis refers to the statistical analysis of written publications, such as books or articles, allowing for the identification of patterns, trends, and emerging themes within a body of literature. This method can offer a macroscopic view of the research landscape, highlighting the evolution of the field, key contributors, and potential future directions.

Our bibliometric study aims to analyze the research landscape on trust in CA over the last fifteen years, highlighting the research trends, patterns, and international relations that characterize this field. This study tries to address the following research questions (RQs) :

1. How has 'Trust in CA' research evolved form 2009 to 2024?
2. How do the number of publications and citations rank among studies, journals/conferences, authors, institutions and funding sponsors?
3. How are the studies regarding CA and trust distributed across different countries and disciplines over the years?
4. What patterns emerge in the analysis of the most used keywords and keyword co-occurance analsis?
5. What patterns emerge in the analysis of co-authorship by countries?
6. What kind of thematic clusters derived from publications?

The remainder of this paper is organized as follows. The following section describes the data collection and methodology. Section 3 presents the performance analysis results and Section 4 presents science mapping analysis results. The last section provides the discussions and implications for future research.

## 2. DATA COLLECTION AND METHODOLOGY

To retrieve essential data for our bibliometric analysis, two important databases, Web of Science (WoS) and Scopus, were examined. While WoS encompasses papers indexed in approximately 12.000 journals, Scopus includes papers indexed in more than 20.000 journals. Additionally, Scopus spans a broader spectrum of disciplines, including science, technology, medicine, social sciences, and humanities. To obtain the most comprehensive dataset for our study, searches were conducted in both WoS and Scopus.

CA may have varied names across disciplines due to their interdisciplinary nature. Additionally, diverse terms arise for CA based on their design and intended use. In order to ensure that the topic of the study was fully captured, we identified all synonyms for 'CA' based on Wang et al. (2023) and integrated them into our search query. Furthermore, alternative terms for 'Trust' were considered, and the search query was formulated as shown in Table 1. To streamline our analysis and focus on the most representative research, we specifically included only journal articles and conference papers, excluding studies not published in English.

**Table 1.** Search query in Scopus and WoS.

| Field tag (Title, abstract, keywords) | TITLE-ABS-KEY ( "chatbot" OR "chat bot" OR "chatterbot" OR "conversational agent" OR "voice assistant" OR "dialogue system" OR "dialog system" OR "conversational assistant" OR "natural language interface" OR "text-based agent " OR "virtual agent" OR "virtual assistant" OR "ChatGPT" OR "GPT" ) AND TITLE-ABS-KEY("trust" OR "trust dimensions" OR "trustworthiness" OR "trustworthy" OR "mistrust" OR "*trust" OR "trust*" ) |
|---|---|
| Document types | Article, Conference Paper |
| Source type | Journal, Conference Proceeding |
| Language | English |
| Time frame | 2009-2024(Q1) |

We searched the title, abstract, and keyword sections of studies published between 2009 and 2024 in the WoS and Scopus databases. The search yielded 691 studies from WoS and 955 studies from Scopus. Due to its extensive coverage of 'trust in CA' records, the data collection process focused exclusively on the Scopus database.

The information retrieved from Scopus includes details such as paper title, author names and affiliations, years, abstract, keywords, references, journal/conference titles, funding/sponsor names, and citation numbers. All collected data was saved in BibTeX and CSV format for further analysis.

We employed two methods that are popular with bibliometric analyses: performance analysis and science mapping. Performance analysis evaluates the contributions of various research elements within a specific field, while science mapping investigates the relationships between these elements (Donthu et al., 2021). To conduct the analysis, we used Microsoft Excel, Scopus, VOSviewer, and Python pyBibX library (Pereira et al., 2023). The results of these two main methods are presented in the following two sections.

## 3. PERFORMANCE ANALYSIS RESULTS

This section presents the performance analysis of 'Trust in CA' topics published between 2009 and March 2024. The latter will be indicated by 2024 Q1 (quarter one) throughout this analysis. Figures and tables present results from the analysis of 955 research publications. In all tables and figures, 'TC' indicates the total number of citations, 'TP' the total number of publications, 'TC/TP' the average number of citations per publication, and 'TC/Y' the average number of citations per year. In most tables, rank is indicated by 'R'.

### 3.1. Descriptive Statistics

Table 2 summarizes the statistical information of 955 studies published between 2009 and April 2024. These studies comprise 57 % articles and 43 % conference papers. The research was conducted by 1.460 institutions from 81 countries. It demonstrates its global reach and multidisciplinary nature. The average collaboration index is 3,83, indicating that papers are typically co-authored by nearly four authors. This is further supported by the low number of single-authored papers and the high number of multi-authored papers. There are in total 3.166 authors contributing an average of 1,15 papers each.

Scopus records contain two types of keywords: author keywords and keywords plus. Author keywords are provided by the original authors, while keywords plus are extracted from the titles of cited references by an automatic algorithm. There are 2.576 author keywords and 4.293 keywords plus, which show the growing vocabulary and changing research topics in 'Trust in CA'

The max h-index is 7, which could mean the field is still developing or that influential publications are yet to emerge. However, the total number of citations (13.224) shows the academic interest and influence of the published research. The average number of documents per institution (2,53) and per year (59,12) indicates sustained and steady contributions to the field, signifying ongoing research interest and continuous development.

**Table 2.** Statistical information of dataset.

| Main Information | Results |
| --- | --- |
| Timespan | 2009-2024(Q1) |
| Total Number of Countries | 81 |
| Total Number of Institutions | 1.460 |
| Total Number of Sources | 543 |
| Total Number of Documents | 955 |
| Total Number of Articles | 541 |
| Total Number of Conference Papers | 414 |
| Average Documents per Author | 1,15 |
| Average Documents per Institution | 2,53 |
| Average Documents per Source | 1,74 |
| Average Documents per Year | 59,12 |
| Total Number of Authors | 3.166 |
| Total Number of Authors Keywords | 2.576 |
| Total Number of Authors Keywords Plus | 4.293 |
| Total Single-Authored Documents | 84 |
| Total Multi-Authored Documents | 871 |
| Average Collaboration Index | 3,83 |
| Max h-index | 7 |
| Total Number of Citations | 13.224 |
| Average Citations per Author | 4,18 |
| Average Citations per Institution | 9,06 |
| Average Citations per Document | 13,98 |
| Average Citations per Source | 24,35 |

Figure 1 shows the distribution of publications by year. As 2024 (Q1) does not represent the entire year, it was not included in this analysis. The number of studies continued to increase from 2017 to 2023. As many studies were published in 2023 as in the years 2009-2022 combined. To analyze if the release of OpenAI's ChatGPT CA might be interlinked with this, we also analyzed the studies focusing on ChatGPT versus the ones focusing on other topics (Fig. 2). In November 2022, OpenAI introduced ChatGPT, a conversational agent similar to a chatbot. Prior to this, the literature encompassed a wide variety of applications and analyses of different conversational agents. However, by 2024 (Q1), for the first time, the number of publications exclusively focusing on ChatGPT exceeded those covering all other conversational agent topics combined (Fig. 2).

Looking at the changes in the number of publications over the years on a source basis, it is observed that journal publications generally followed a parallel trend. Conference papers, on the other hand, showed a declining trend from 2017 to 2020 but began to increase again as of 2021, reaching its so-far peak in 2023.

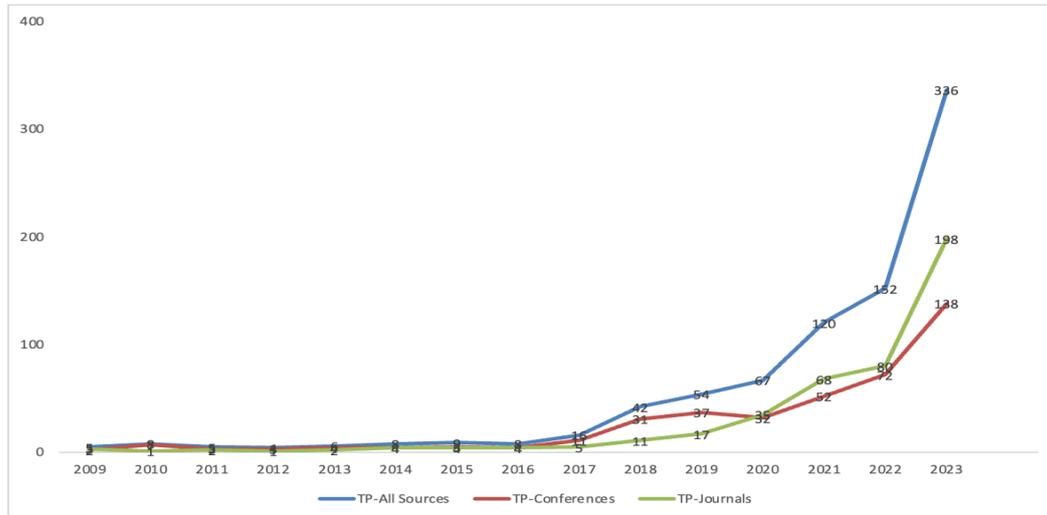

**Figure 1.** Number of publications by year.

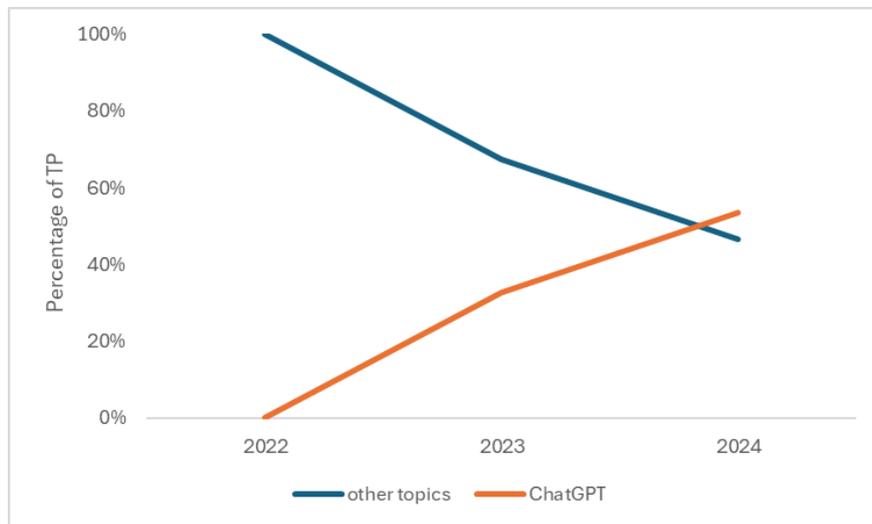

**Figure 2.** Yearly percentage of publications focused on ChatGPT vs. other CA.

### 3.2. Leading Journals and Conferences

This section provides lists of the top ten leading journals and conferences in 'Trust in CA' research. Table 3 ranks the top ten journals by total publication number and includes data on total citations, citations per publication, impact factor, and h-index. The International Journal of Human-Computer Interaction is the most active journal in this field, with a total of 18 publications. However, the International Journal of Human-Computer Studies has the highest number of citations, indicating that research published in this journal has received significant attention and recognition within the field. Despite their contributions, these journals are behind others in impact factor and h-index. Nevertheless, these metrics are important for gauging the prestige of journals.

**Table 3.** Top 10 journals publishing papers on 'Trust in CA'.

| R | Journal Name | TP | TC | TC/TP | Impact Factor | h-index |
|---|---|---|---|---|---|---|
| 1 | International Journal of Human-Computer Interaction | 18 | 318 | 17,67 | 4,787 | 76 |
| 2 | Journal of Medical Internet Research | 17 | 494 | 29,06 | 7,517 | 158 |
| 3 | Computers in Human Behavior | 14 | 405 | 28,93 | 9,79 | 203 |
| 4 | Proceedings of the ACM on Human-Computer Interaction | 13 | 583 | 44,85 | 4,568 | 38 |
| 5 | Frontiers in Psychology | 9 | 48 | 5,33 | 3,884 | 133 |
| 6 | International Journal of Human Computer Studies | 8 | 646 | 80,75 | 6,381 | 129 |
| 7 | Journal of Business Research | 8 | 302 | 37,75 | 11,063 | 217 |
| 8 | Journal of Retailing and Consumer Services | 8 | 50 | 6,25 | 11,103 | 104 |
| 9 | Psychology and Marketing | 7 | 382 | 54,57 | 4,562 | 124 |
| 10 | JMIR Human Factors | 7 | 28 | 4,00 | 2,506 | 22 |

Table 4 shows the top ten conferences which are publishing most frequently in the area of 'Trust in CA'. The 'Conference on Human Factors in Computing Systems' (CHI) is in the lead with a total of 30 publications and 816 citations, demonstrating its significant impact in the field. It is followed by the International Conference on Intelligent Virtual Agents (IVA) and the Conference on Empirical Methods in Natural Language Processing (EMNLP). Both have 20 publications, but with large differences in citations, showing the greater impact of IVA.

**Table 4.** Top 10 conferences publishing papers on 'Trust in CA'.

| R | Conference Name | Abbreviation | TP | TC | TC/TP |
|---|---|---|---|---|---|
| 1 | Conference on Human Factors in Computing Systems | CHI | 30 | 816 | 27,20 |
| 2 | International Conference on Intelligent Virtual Agents | IVA | 20 | 257 | 12,85 |
| 3 | Conference on Empirical Methods in Natural Language Processing | EMNLP | 20 | 36 | 1,80 |
| 4 | Annual Hawaii International Conference on System Sciences | HICSS | 18 | 162 | 9,00 |
| 5 | CEUR Workshop | CEUR-WS | 14 | 3 | 0,21 |
| 6 | International Conference on Human-Robot Interaction | HRI | 13 | 64 | 4,92 |
| 7 | International Conference on Human-Agent Interaction | HAI | 7 | 71 | 10,14 |
| 8 | Annual Meeting of the Association for Computational Linguistics | ACL | 6 | 41 | 6,83 |
| 9 | International Conference on Intelligent User Interfaces | IUI | 5 | 65 | 13,00 |
| 10 | Annual Conference of the International Speech Communication Association | ISCA Speech | 5 | 19 | 3,80 |

## 3.3. Top 20 Highly Influential Papers

To assess the most significant studies in the field of 'Trust in CA', we looked at the total number of citations for each publication. The total number of citations serves as a measure of the study's impact and significance within the research field. Table 5 lists the top 20 most-cited papers in this domain. This table includes the title of each paper, year of publication, authors, source, number of citations, and average number of citations per year. The paper drawing the greatest interest, 'Alexa, are you listening? Privacy perceptions, concerns and privacy-seeking behaviors with smart speakers' by Lau et al. (2018), has been published in the Proceedings of the ACM on Human-Computer Interaction. It has garnered attention with an average of 66 citations per year, totalling 397 citations. This work sheds light on user adoption processes, as well as their perceptions of privacy and trust, and concerns regarding smart speakers.

Among the top 20 most influential articles listed, there are ten articles published in 2020 and 2021, with a total of 1713 citations. This indicates that these papers have achieved high citation counts in a short period and have made significant contributions to their field. All of the top 20 most cited articles have more than 100 citations. Furthermore, three of the top 20 most cited papers were published in the 'International Journal of Human Computer Studies' and two in the 'Conference on Human Factors in Computing Systems'.

**Table 5.** Top 20 highly cited papers on 'Trust in CA'.

| R | Title | Year | Authors | Source | TC | TC/Y |
|---|-------|------|---------|--------|-----|------|
| 1 | Alexa, are you listening? Privacy perceptions, concerns, and privacy-seeking behaviors with smart speakers | 2018 | Lau J.; Zimmerman B.; Schaub F. | Proceedings of the ACM on Human-Computer Interaction | 397 | 66 |
| 2 | Almost human: Anthropomorphism increases trust resilience in cognitive agents | 2016 | de Visser E.J.; Monfort S.S.; McKendrick R.; Smith M.A.B.; McKnight P.E.; Krueger F.; Parasuraman R. | Journal of Experimental Psychology: Applied | 311 | 39 |
| 3 | Acceptability of artificial intelligence (AI)-led chatbot services in healthcare: A mixed-methods study | 2019 | Nadarzynski T.; Miles O.; Cowie A.; Ridge D. | Digital Health | 297 | 59 |
| 4 | Adoption of AI-based chatbots for hospitality and tourism | 2020 | Pillai R.; Sivathanu B. | International Journal of Contemporary Hospitality Management | 283 | 71 |
| 5 | What makes a good conversation? Challenges in designing truly conversational agents | 2019 | Clark L.; Pantidi N.; Cooney O.; Doyle P.; Garaialde D.; Edwards J.; Spillane B.; Gilmartin E.; Murad C.; Munteanu C.; Wade V.; Cowan B.R. | Conference on Human Factors in Computing Systems - Proceedings | 258 | 52 |
| 6 | Understanding the attitude and intention to use smartphone chatbots for shopping | 2020 | Kasilingam D.L. | Technology in Society | 256 | 64 |
| 7 | Enhancing user experience with conversational agent for movie recommendation: Effects of self-disclosure and reciprocity | 2017 | Lee S.; Choi J. | International Journal of Human Computer Studies | 206 | 29 |
| 8 | The human side of human-chatbot interaction: A systematic literature review of ten years of research on text-based chatbots | 2021 | Rapp A.; Curti L.; Boldi A. | International Journal of Human Computer Studies | 205 | 68 |
| 9 | "Hey Google is it ok if I eat you?" Initial explorations in child-agent interaction | 2017 | Druga S.; Breazeal C.; Williams R.; Resnick M. | IDC 2017 - Proceedings of the 2017 ACM Conference on Interaction Design and Children | 199 | 28 |
| 10 | Challenges in Building Intelligent Open-domain Dialog Systems | 2020 | Huang M.; Zhu X.; Gao J. | ACM Transactions on Information Systems | 187 | 47 |
| 11 | Alexa, she's not human but… Unveiling the drivers of consumers' trust in voice-based artificial intelligence | 2021 | Pitardi V.; Marriott H.R. | Psychology and Marketing | 164 | 55 |
| 12 | In bot we trust: A new methodology of chatbot performance measures | 2019 | Przegalinska A.; Ciechanowski L.; Stroz A.; Gloor P.; Mazurek G. | Business Horizons | 160 | 32 |
| 13 | My Chatbot Companion - a Study of Human-Chatbot Relationships | 2021 | Skjuve M.; Følstad A.; Fostervold K.I.; Brandtzaeg P.B. | International Journal of Human Computer Studies | 146 | 49 |
| 14 | Response to a relational agent by hospital patients with depressive symptoms | 2010 | Bickmore T.W.; Mitchell S.E.; Jack B.W.; Paasche-Orlow M.K.; Pfeifer L.M.; O'Donnell J. | Interacting with Computers | 132 | 9 |
| 15 | Alexa, do voice assistants influence consumer brand engagement? – Examining the role of AI powered voice assistants in influencing consumer brand engagement | 2021 | McLean G.; Osei-Frimpong K.; Barhorst J. | Journal of Business Research | 129 | 43 |
| 16 | Machine heuristic: When we trust computers more than humans with our personal information | 2019 | Sundar S.; Kim J. | Conference on Human Factors in Computing Systems - Proceedings | 126 | 25 |
| 17 | An experimental study of public trust in AI chatbots in the public sector | 2020 | Aoki N. | Government Information Quarterly | 117 | 29 |
| 18 | User experiences of social support from companion chatbots in everyday contexts: Thematic analysis | 2020 | Ta V.; Griffith C.; Boatfield C.; Wang X.; Civitello M.; Bader H.; DeCero E.; Loggarakis A. | Journal of Medical Internet Research | 114 | 29 |
| 19 | "In A.I. we trust?" The effects of parasocial interaction and technopian versus luddite ideological views on chatbot-based customer relationship management in the emerging "feeling economy" | 2021 | Youn S.; Jin S.V. | Computers in Human Behavior | 112 | 37 |
| 20 | Does a Digital Assistant Need a Body? the Influence of Visual Embodiment and Social Behavior on the Perception of Intelligent Virtual Agents in AR | 2018 | Kim K.; Boelling L.; Haesler S.; Bailenson J.; Bruder G.; Welch G.F. | Proceedings of the 2018 IEEE International Symposium on Mixed and Augmented Reality, ISMAR 2018 | 111 | 19 |

## 3.4. Most Productive and Influential Authors

This section presents an analysis of the most productive and influential authors in the Trust in CA' field. Table 6 shows the top 10 authors who have the highest number of publications. In instances where authors have the same number of publications, their rank is determined by their total citation number. Additionally, Table 6 provides the h-index for these authors in both the specific field of 'Trust in CA' and a more general context. The h-index is a metric that reflects an author's productivity and impact within the scholarly community. An author's h-index is determined by the quantity of his or her works that have been cited at least 'h' times. For example, an h-index of 5 means the author has at least 5 publications that have each been cited 5 times or more.

Table 6 shows that Wolfgang Minker has the highest contribution to the field with 11 publications. He is followed by Deborah Richards with 9, and Tetsuya Matsui with 8 publications. According to the total number of citations, Timothy Bickmore stands out as the most cited author, having 7 publications cited 238 times in total. The highest field-specific h-index belongs to Deborah Richards, indicating that she has a substantial number of publications that are frequently cited by other researchers in the field.

**Table 6.** Top 10 most productive and influential authors in 'Trust in CA'.

| R | Author | TP | TC | TC/TP | h-index (Field) | h-index (General) | Country |
|---|---|---|---|---|---|---|---|
| 1 | Wolfgang Minker | 11 | 91 | 8,27 | 5 | 37 | Germany |
| 2 | Deborah Richards | 9 | 102 | 11,33 | 7 | 41 | Australia |
| 3 | Tetsuya Matsui | 8 | 27 | 3,38 | 3 | 6 | Japan |
| 4 | Timothy Bickmore | 7 | 238 | 34,00 | 5 | 73 | USA |
| 5 | Tze Wei Liew | 7 | 177 | 25,29 | 5 | 16 | Malaysia |
| 6 | Su-Mae Tan | 7 | 177 | 25,29 | 5 | 13 | Malaysia |
| 7 | Matthias Kraus | 7 | 72 | 10,29 | 3 | 16 | Germany |
| 8 | Jean-Arthur Micoulaud-Franchi | 6 | 105 | 17,50 | 4 | 42 | France |
| 9 | Piere Philip | 6 | 105 | 17,50 | 4 | 60 | France |
| 10 | Nicolas Wagner | 6 | 72 | 12,00 | 3 | 6 | Germany |

## 3.5. Affiliation and Funding Sponsor

We examined the affiliation institutions of authors and determined the top ten institutions according to total publications in the 'Trust in CA' research area. Table 7 presents the top ten affiliation institutions. Northeastern University in USA ranks first with the highest number of publications and most citations, followed by the Centre National de la Recherche Scientifique (CNRS) in France and the University of Florida in the USA. Two German universities, the Universität Ulm and the Universität Duisburg-Essen, are the top-ranking institutions in terms of the number of publications.

**Table 7.** Top 10 affiliation institutions in 'Trust in CA'.

| R | Affiliation Instituations | Country | TP | TC | TC/TP |
|---|---|---|---|---|---|
| 1 | Northeastern University | USA | 15 | 352 | 23,47 |
| 2 | Centre National de la Recherche Scientifique (CNRS) | France | 13 | 206 | 15,85 |
| 3 | University of Florida | USA | 13 | 96 | 7,38 |
| 4 | Universität Ulm | Germany | 13 | 94 | 7,23 |
| 5 | Universität Duisburg-Essen | Germany | 12 | 93 | 7,75 |
| 6 | Symbiosis International Deemed University | India | 11 | 140 | 12,73 |
| 7 | Macquarie University | Australia | 11 | 122 | 11,09 |
| 8 | Delft University of Technology | Netherlands | 11 | 89 | 8,09 |
| 9 | University of Washington | USA | 11 | 73 | 6,64 |
| 10 | Universität Zürich | Switzerland | 9 | 133 | 14,78 |

To understand the role of global funding bodies in advancing 'Trust in CA' research, we reviewed the sponsors behind relevant publications. Table 8 displays the leading 10 funding sponsors in this domain. The National Natural Science Foundation of China, the USA National Science Foundation, and the European Union (EU) Horizon 2020 emerged as the top supporters. This funding landscape underscores the dominant roles of the USA, EU, and China.

**Table 8.** Top 10 funding sponsors in 'Trust in CA'.

| R | Funding Sponsor | Country | TP | TC | TC/TP |
|---|---|---|---|---|---|
| 1 | National Natural Science Foundation of China | China | 39 | 424 | 10,87 |
| 2 | National Science Foundation | USA | 31 | 753 | 24,29 |
| 3 | Horizon 2020 | EU | 30 | 280 | 9,33 |
| 4 | Deutsche Forschungsgemeinschaft | Germany | 15 | 237 | 15,80 |
| 5 | Engineering and Physical Sciences Research Council | UK | 14 | 147 | 10,50 |
| 6 | National Institutes of Health | USA | 13 | 135 | 10,38 |
| 7 | European Commission | EU | 12 | 108 | 9,00 |
| 8 | Japan Society for the Promotion of Science | Japan | 11 | 134 | 12,18 |
| 9 | Fundamental Research Funds for the Central Universities | China | 10 | 195 | 19,50 |
| 10 | Air Force Office of Scientific Research | USA | 9 | 474 | 52,67 |

### 3.6. Discipline Analysis

To analyze the number of publications from each discipline, we utilized the Scopus analysis tool. Table 9 presents the number of publications by discipline from 2009 to 2024 (Q1), highlighting the top ten disciplines. The majority of publications originate from the field of computer science, followed by social sciences and engineering. Note that a single study may impact multiple disciplines (e.g., both computer science and engineering), so the total number of publications exceeds 955.

**Table 9.** Top 10 disciplines in 'Trust in CA'.

| R | Discipline | TP |
|---|---|---|
| 1 | Computer Science | 633 |
| 2 | Social Sciences | 252 |
| 3 | Engineering | 193 |
| 4 | Business, Management and Accounting | 133 |
| 5 | Medicine | 107 |
| 6 | Decision sciences | 79 |
| 7 | Psychology | 74 |
| 8 | Arts and Humanities | 66 |
| 9 | Mathematics | 57 |
| 10 | Economics, Econometrics and Finance | 26 |
|  | Total | 1620 |

Figure 3 displays the publication output of the top ten disciplines from 2017 to 2024 (Q1), presented in total numbers (left) and percentage share of all publications (right). While computer science remains the leading discipline in terms of publication number, its percentage share is declining. This trend is attributed to the rising number of publications from other disciplines, such as social sciences.

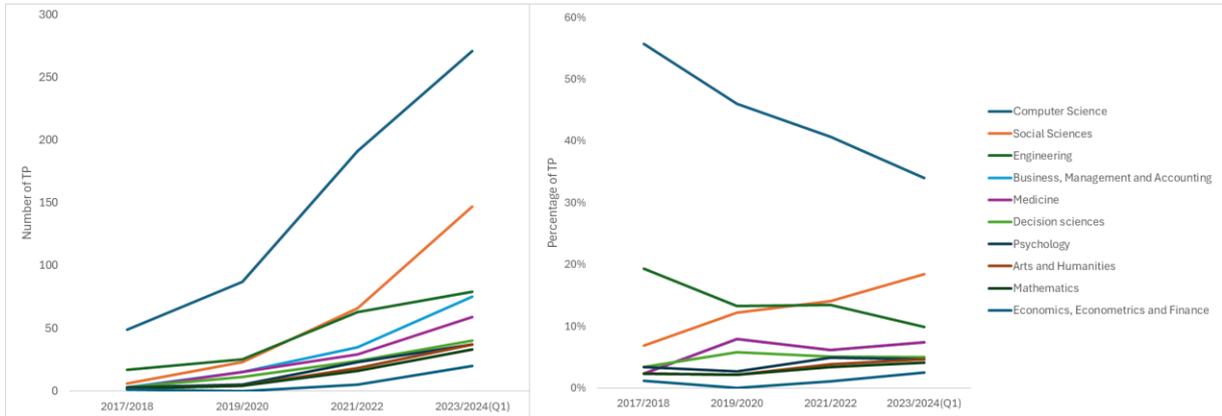

**Figure 3.** Total numbers and percentage of publications by discipline.

### 3.7. Country Analysis

Figure 4 shows the distribution of publications number by country worldwide from 2009 to 2024 (Q1). The USA leads significantly with a total of 299 publications. Although Germany holds the second position overall with 122 publications, its publication numbers have declined since 2021/2022 (Fig. 5). In 2023/2024 (Q1), China surpassed all other countries except the USA in the number of publications.

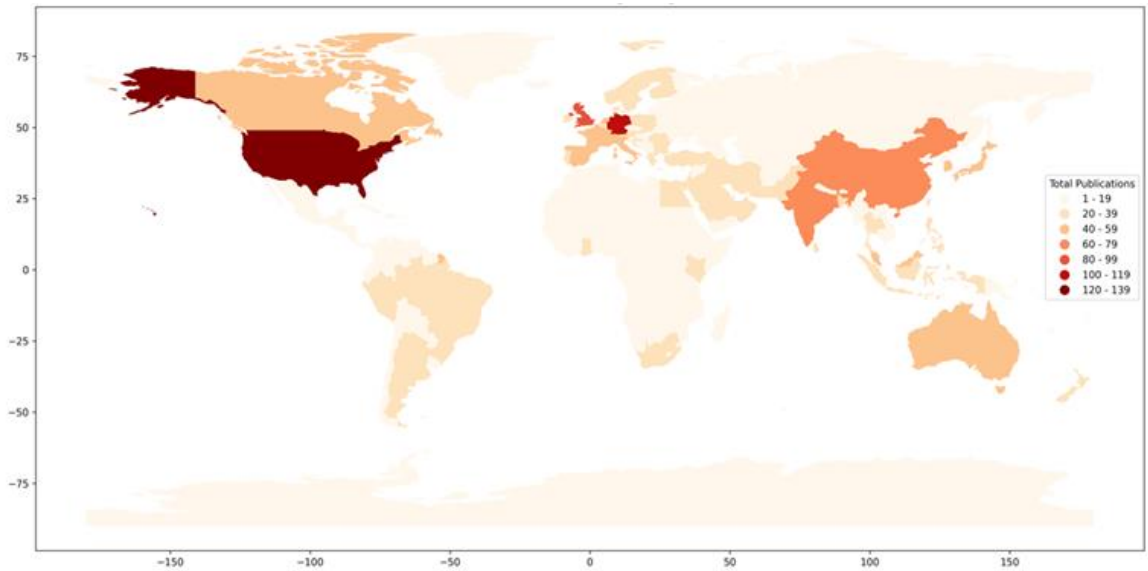

**Figure 4.** Number of publications by country.

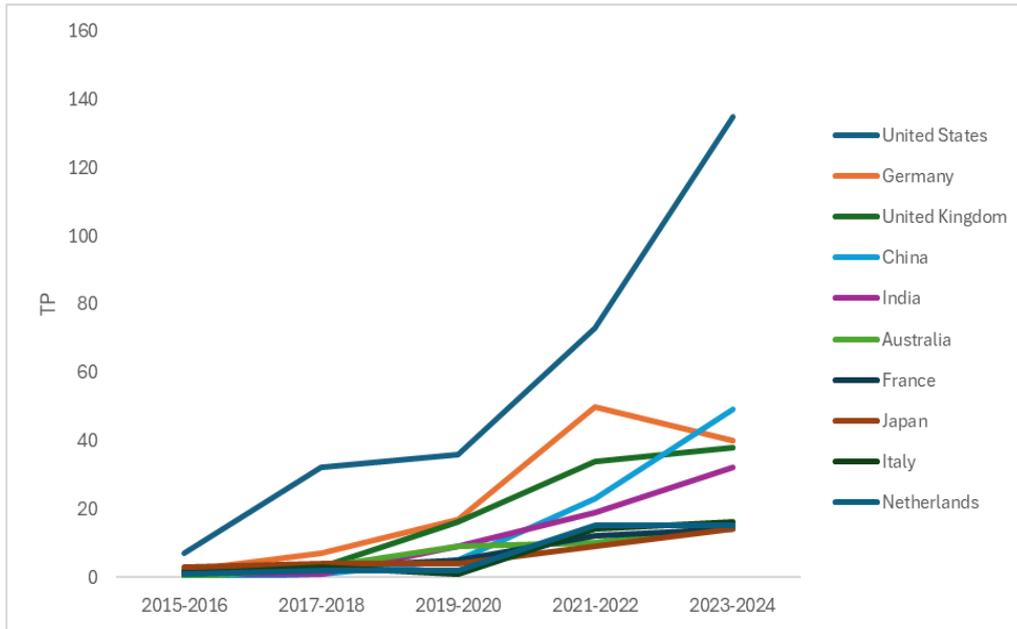

**Figure 5.** The top 10 countries by total publications, 2015-2024 (Q1).

### 3.8. Keyword Analysis

We identified the most frequently used keywords for each year from 2019 to 2024 (Q1) and analyzed their percentage share within the top five keywords over these years. Figure 6 shows that the three keywords 'trust', 'artificial intelligence', and 'chatbot(s)' were predominant the whole time. 'Anthropomorphism' disappeared from the top five list after 2019, just like 'conversational agent(s)' after 2022. 'ChatGPT' is a new predominant keyword since 2023.

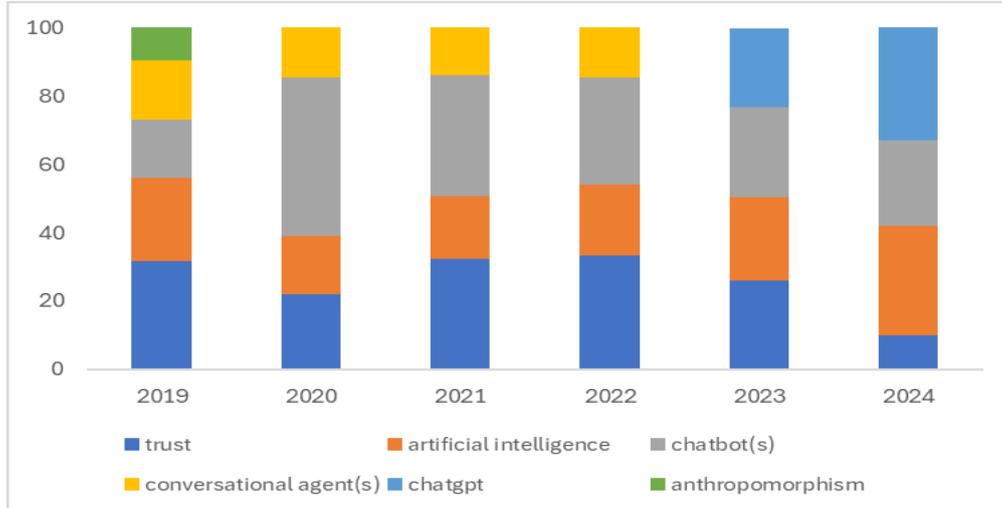

**Figure 6.** The top five keywords per year and their percentage share from 2019-2024 (Q1).

Figure 7 provides a ranking of the fifteen most frequent three-word phrases based on their occurrence. The term 'human-computer interaction' emerges as the most frequent, indicating that research on 'Trust in CA' predominantly focuses on human aspects rather than technological developments.

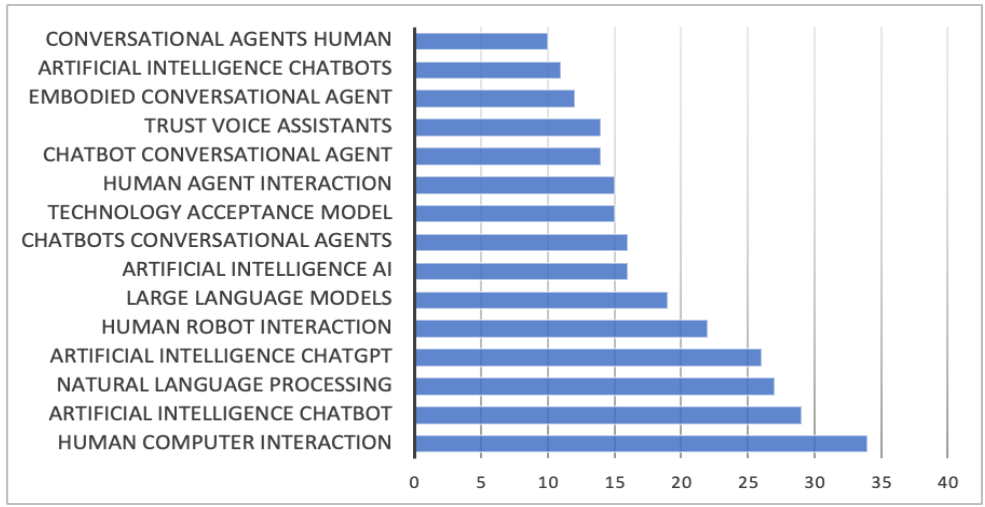

**Figure 7.** Top 15 three-word phrases.

## 4. SCIENCE MAPPING ANALYSIS RESULTS

This section presents the findings from the science mapping analysis of 955 studies on 'Trust in CA'. We conducted keyword co-occurrence analysis, citation analysis, and bibliographic coupling analysis, using VOSviewer software.

### 4.1. Keyword Co-occurrence Analysis

Keywords co-occurrence analysis identifies the frequency of commonly used keywords, revealing the relationships and interactions between various dimensions. For mapping, we selected 498 keywords that appear at least five times from a pool of 8.697 words. This resulted in 7.064 connections and 14 clusters. Figure 8 displays the network of keywords that authors use most commonly in their publications. Larger circles represent more frequent keyword occurrences, and the colors of different circles indicate distinct clusters. 'Artificial intelligence' is the leading keyword with 618 occurrences, exhibiting the strongest connectivity, evidenced by 1676 links to other keywords. It is followed by 'chatbot' (548), 'ChatGPT' (447), 'chatbots' (286), 'conversational agents' (262), and 'trust' (202). The clusters shown in different colors represent the most frequently used keywords together. For example, in the cluster colored purple, the terms 'Large language models', 'natural language processing', and 'generative AI' have frequently occurred together.

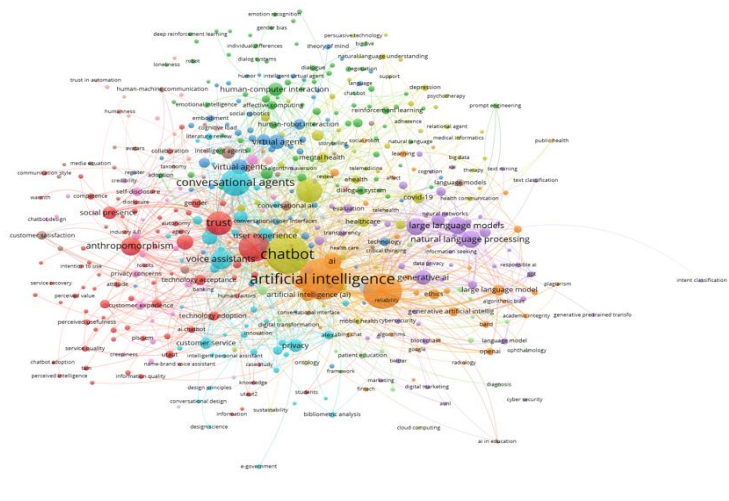

**Figure 8.** Keyword co-occurrence network.

Figure 9 illustrates how the occurrence of keywords has changed over the past five years. The size of the circle is indicative of a topic's prevalence, with larger circles representing a more frequently discussed subject, while yellow circles denote the most popular topics. It is noteworthy that popular topics include ChatGPT, large language models, generative AI, and GPT.

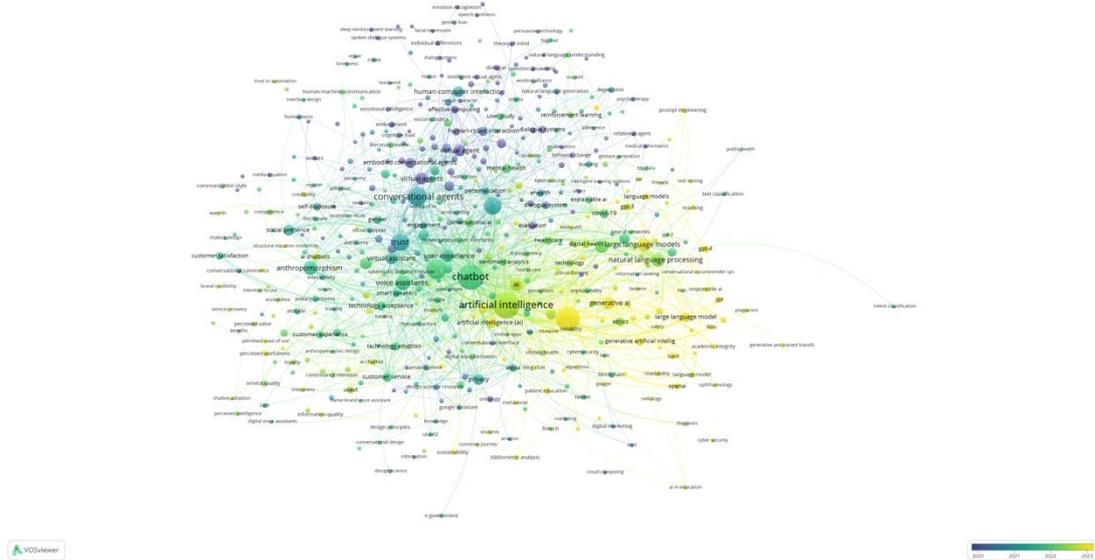

**Figure 9.** Keyword co-occurrence network over the years.

### 4.2. Co-authors Analysis by Countries

For the co-authors analysis by countries, we used VOSviewer to show from which countries authors collaborated on publications. We focused on countries with a minimum of ten publications. From 96 countries in total, we had an output of 28 countries fulfilling this threshold. In total, there were 96 links between the countries with an output of 347 joint publications (Fig. 10). The USA has the most links (25) and publications based on international collaborations (117), followed by the UK (23 links, 75 publications), and Germany (18 links, 54 publications). The most joint publications have the USA and UK (18), the USA and China (16), and the USA and Germany (9).

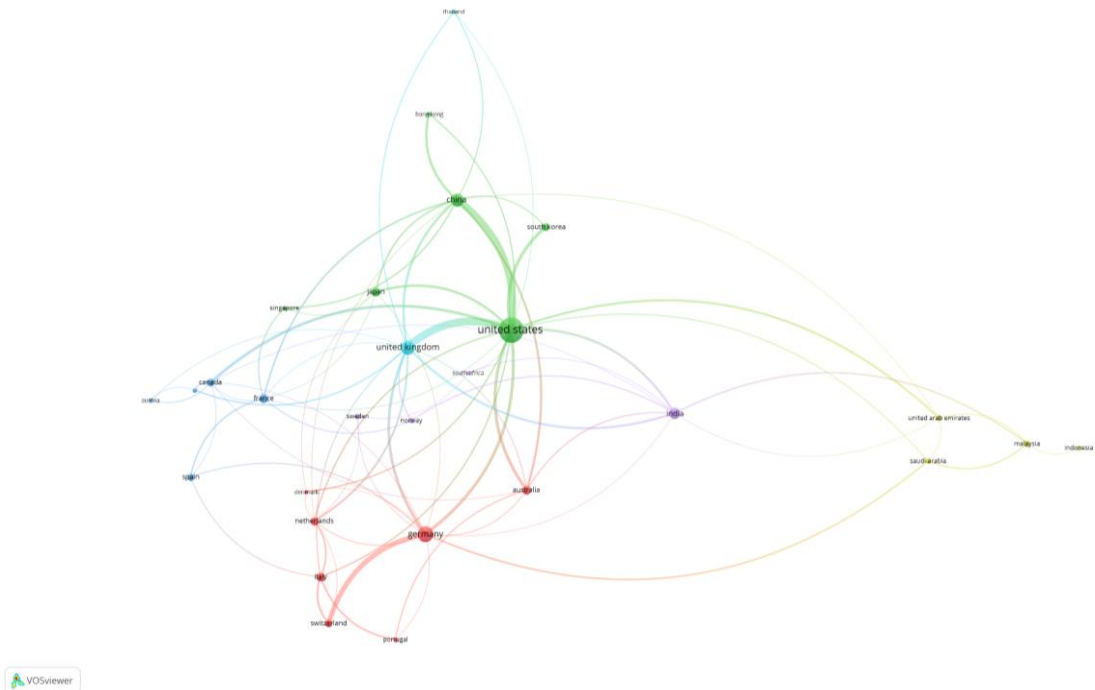

**Figure 10.** Co-authorship analysis by country.

## 4.3. Co-citation Analysis by Sources

Co-citation analyses identify the frequency and connections of shared citations among different elements such as articles, authors, sources, and countries. A co-citation relationship is established when two elements are simultaneously cited by a third element. The strength of this relationship increases as the frequency of co-citations increases, indicating a semantic link. In this study, we examined the co-citation network of sources to identify the most influential journals and conference proceedings.

For the co-citation analysis by source, we used VOSviewer and filtered for a minimum of 100 citations per source, which led to an output of 33 items (Fig. 11). The most co-citation links have the Journal of Computers in Human Behavior (32 links, 930 citations) and the Management Information Systems Quarterly (32 links, 415 citations), followed by the International Journal of Human-Computer Interaction (31 links, 332 citations), and the Journal of Business Research (30 links, 447 citations).

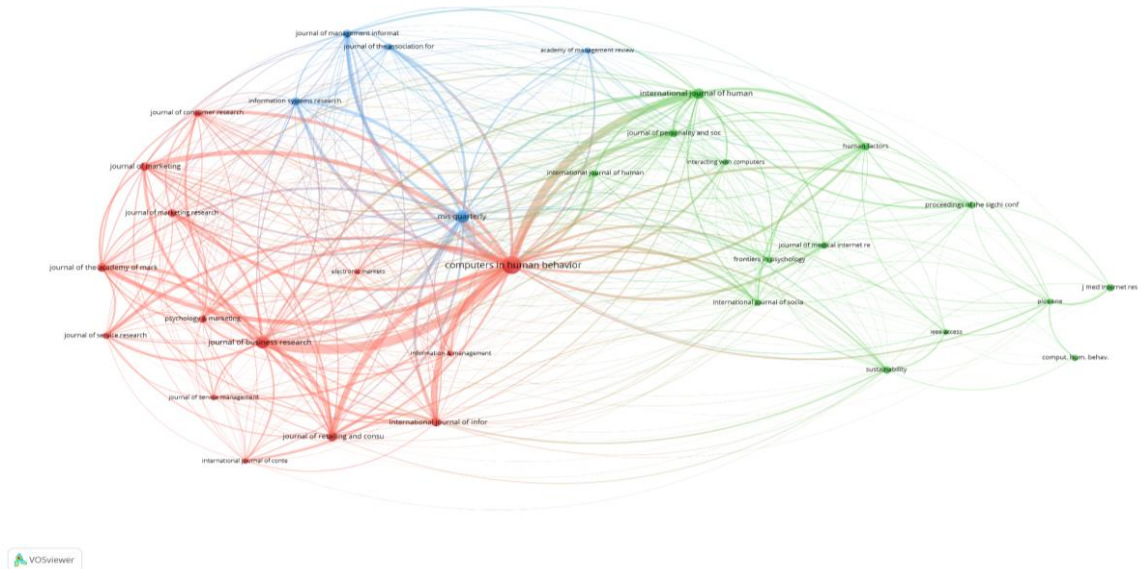

**Figure 11.** Co-citation analysis by source.

## 4.4. Clustering Analysis

To identify the main themes present in the publications and to understand their distribution, we analyzed the 955 publications using text clustering methods. These methods group data points into distinct clusters based on content similarities and differences.

In this study, we created a corpus consisting of the titles and abstracts of publications. Clustering requires a numerical vector representation of textual features. Before constructing vector representations, we applied some preprocessing steps to improve representation quality: correcting typos, removing extra spaces, converting uppercase to lowercase, removing email addresses and URLs, removing punctuation, and cleaning English stopwords. We transformed the text data into high-dimensional numerical vectors using the BERT word embedding technique (Devlin et al., 2019), which captures semantic information more effectively than traditional methods like TF-IDF (Kalyan et al., 2021). Using these numerical vectors, we applied a k-means clustering algorithm to group the publications. A key challenge in k-means is determining the optimal number of clusters (k), so we tested scenarios with k=5, k=6, and k=7. Expert judgment is often needed to evaluate the resulting clusters. After a detailed examination, we found that six clusters were the most appropriate.

To visualize and analyze the distribution of publications within these clusters, we reduced the high-dimensional vectors to a two-dimensional space using t-distributed stochastic neighbor embedding (t-SNE). Figure 12 illustrates the results, where each circle represents a publication, and the six colors denote different clusters.

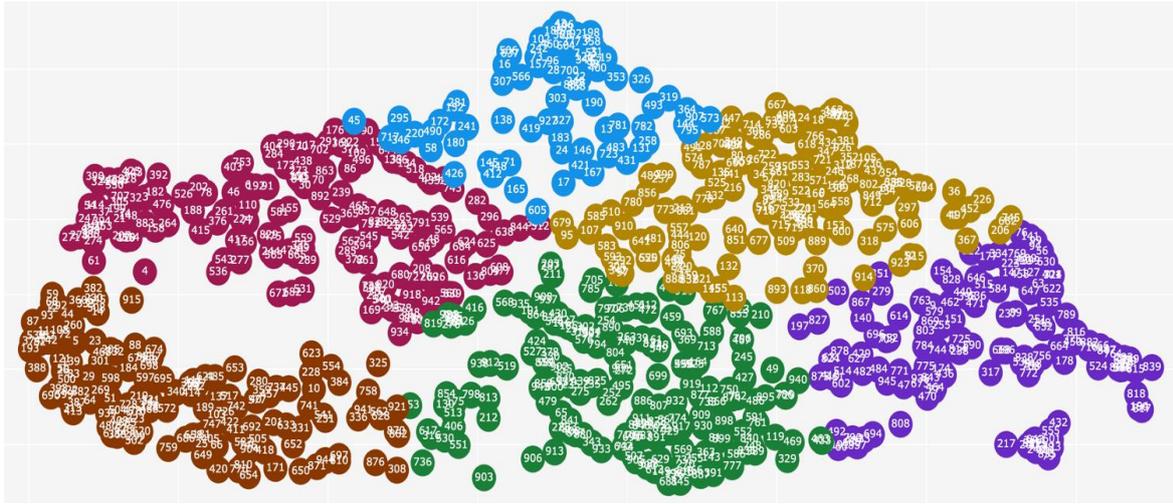

**Figure 12.** Clusters of publications.

We analyzed the research focus areas of these clusters to identify the major themes and assigned labels to each cluster based on these themes. This analysis provided a clearer picture of the main research interest within the field. Table 10 shows the clusters, their colors from Figure 12, the number of publications in each cluster, labels, and each cluster's research focus. Unified by the overarching theme of trust, each cluster addresses a specific application area or aspect of CA.

**Table 10.** Proposed clusters with their theme.

| Number | Color | TP in Cluster | Label | Research Focus |
|---|---|---|---|---|
| Cluster 1 | Purple | 149 | Voice Assistants: Trust, Privacy, User Interaction | This cluster predominantly focuses on voice assistants, such as Alexa. The main themes revolve around trust and privacy concerns associated with these technologies, as well as user interaction and experience. |
| Cluster 2 | Pink | 179 | Human-LLMs interaction and trust | This cluster deals with large language models (LLMs) and their applications, including ChatGPT. The focus is on the trust in these AI systems, the data they use, and their interactions with humans. |
| Cluster 3 | Green | 202 | Trust and Social Interaction in CA | This cluster emphasizes trust in CA, particularly virtual agents. It includes discussions on social interactions with these agents and explores embodied CA. |
| Cluster 4 | Yellow | 172 | CA in Customer Service | This cluster focuses on chatbots used in customer service, this cluster explores the trust issues and implementation challenges of CA in customer-facing applications. |

| Cluster 5 | Brown | 158 | CA in Health | This cluster deals with the use of CA in health and medical contexts. It highlights the trust and reliability of health information provided by these technologies. |
| Cluster 6 | Blue | 86 | CA in Education | This cluster focuses on the adoption of CA in educational settings. It examines trust and the perceived impact of CA on students and educators. |

## 5. DISCUSSION AND IMPLICATIONS FOR FUTURE RESEARCH

In this study, we investigated the predominant trends in the thematic field of CA and trust and how the publications landscape has developed and changed over the last 15 years from 2009 to 2024 (Q1). In the following, we will discuss the results and give implications for future research based on the data on the analyzed data.

After steady publication numbers were measured in the first few years of the period under review, these have risen significantly since 2016, with a particular surge since 2022. The rapidly rising publication numbers clearly show that the topic of 'trust in CA' is attracting increasing interest. One reason for this - especially for the peaking numbers since 2022 - could be the release of ChatGPT by OpenAI in 2022. ChatGPT quickly became popular, attracting a rapidly growing user base (Ma et al., 2024) and shifting the focus of research in the field of CA and trust. ChatGPT has advanced speech understanding and generation capabilities that enable natural user interaction in a user-friendly chat-based interface known by many people from other chat applications from e.g. social media. ChatGPT's extensive training on various datasets provides a broad knowledge of different topics and might be one reason for its popularity. We believe that, based on the numbers of our analysis, ChatGPT had an impact on two levels: (1) the application as a research object itself: ChatGPTs popularity shows that it might be more accessible than other CA for people without a computer science background and therefore also attracts researchers' attention as described previously. Researchers can use ChatGPT to conduct studies e.g. its trustworthiness in field-specific medical information output (e.g. Aguiar de Sousa et al., 2024; Ebrahimi et al., 2023). (2) The application as a research tool in human-AI interaction: ChatGPT is the first CA that has been used by thousands of people all over the world making it an important tool to be analyzed in terms of human-technology interaction e.g. peoples' perceived trustworthiness and their attitude towards CA applications (e.g. Bodani et al., 2023; Maheshwari, 2023).

These hypotheses are based on the increasing number of publications solely focusing on ChatGPT rather than a broad understanding of CA or other CA applications. In 2024 (Q1), for the first time, there were more publications on trust in ChatGPT than on any other trust and CA-related topic. Although it was released only one year before, it was already the third most co-occurring keyword after 'artificial intelligence' and 'trust' in 2023. In 2024 (Q1), it was even the predominant keyword. The diversification of disciplines publishing in the thematic field of CA and trust is an indicator that AI has become more approachable to people without coding skills. ChatGPT is one example of how user interfaces are more accessible than earlier CA applications. Since our study revealed a total of 3.166 authors contributing only an average of 1,15 papers each, the field seems to attract a diverse range of new-to-the-field researchers and ideas that do not solely focus on trust and CA in their research work.

ChatGPT might be one reason why researchers from other disciplines include CA in their trust-related research. On the one hand, they can use a generative AI chatbot without a computer science background and integrate it into their studies. On the other hand, due to the large number of ChatGPT users, it becomes more important for other disciplines to investigate its relation with their field (e.g. how people use and trust it as an information resource regarding different topics and how accurate its answers are). However, it is also noticeable that some of the most cited articles focus on another application: Alexa from Amazon. This could have been the starting point for other disciplines focusing on human-CA and trust research because 2017/2018 is the first time a real difference in research discipline has been visible.

Remarkably, 19 out of 20 most-cited articles are from 2016-2021. Only one paper is from 2010. The predominant years are 2020 and 2021, indicating substantial recent research and citation activity.

Analyzing the first quarter of 2024 within this data set was challenging regarding absolute numbers because (of course) TP, TC, keywords, etc. seemed to decline. However, the percentage share of keywords, disciplines, and country distributions, for example, can provide a good insight into how the research area may change within months and in the coming years. ChatGPT clearly impacts the publication field based on keyword occurrence and publications focusing solely on ChatGPT. *It would be interesting to explore which disciplines focus more on ChatGPT and which prioritize other conversational AI topics in future studies. With the release of more ChatGPT-like free-to-use applications, it will be interesting to observe how the publication landscape evolves in the coming years and whether ChatGPT will maintain its popularity in scientific 'Trust in CA' research.*

Overall, there is a human-centered approach to investigating trust in CA. More studies focus on user perception than on developing trustworthy technology. This is particularly evident in the words used in the titles and abstracts of all publications for the entire period analyzed: the most used single word is 'human' way before 'systems' and 'virtual'. Also, the three-word analysis shows human-centered research in the field: the word combination 'human computer interaction' dominates before 'artificial intelligence chatbot' and 'natural language processing'. Within the ten most dominating disciplines, there are several that usually focus more on humans than on technological applications, like social sciences, medicine, psychology, and humanities. But, also for other disciplines like computer sciences, engineering, and economics, investigating human trust in CA is very important as their dominance in the research field shows. Until 2024 (Q1), computer sciences has been by far the predominant discipline for 'Trust in CA' research, and its numbers of TP are rising, but its percentage share of total publications has been decreasing for years. From 2017/2018 with 56%, it decreased to 34% in 2023/2024 (Q1). The engineering sciences fell from 19 % to 10% during this period. Within these eight years, TP in the social sciences rose from 7% in 2017/2018 to 18 % in 2023/2024 (Q1) and took second place. In third place is Engineering when it comes to total paper numbers, but Medicine and Business/Management show bigger growth in the analyzed eight year period. . It is notable that the total number of papers by discipline exceeds the analyzed number of 955 by far with 1620. This shows that many papers are already interdisciplinary. *We assume that interdisciplinarity will be even more important in the upcoming years in researching 'Trust and CA'. For future research, we assume that computer science will stay the most important discipline when it comes to 'Trust in CA' research. Mainly because computer scientists are the leading group of researchers responsible for the advancement of CA and trustworthy AI. The three disciplines of social science, medicine, and business/management could be of special interest for CA applications and trust research due to their significant increase in this field in recent years. However, engineering is in third place after computer science and social science so far and, therefore, should not be underestimated in its importance for trust in CA research.*

The cluster analysis confirms these findings and shows in which sectors the most research on 'Trust in CA' has been conducted. In addition to computer science and psychology-led topics such as 'human-to-human interaction and trust' or 'voice assistants', we could identify areas such as social interaction, customer service, health, and education. *Due to the rising diversification when it comes to disciplines focusing on 'Trust in CA' research, we believe that these topics will gain even more relevance in the upcoming years, leading to the need for more interdisciplinary research. It seems helpful to conduct a systematic literature review of publications from recent years to gain insight into the different disciplines and their contributions to the field regarding research methodologies, specific topics, and CA applications.*

When it comes to the research activity of different countries, the USA inherits a predominant role in Trust in CA' research. Researchers based in the US publish by far the most research articles and conference papers related to CA and trust. One reason for this might be the 'big player' companies that are located in the USA like Microsoft, IBM, Meta, Google, and Amazon, that have their own research departments like Google DeepMind, Microsoft Research, Meta Research, and IBM Research. These research centers focus on AI research and might have more money available to conduct cost-, time-, and personnel-intensive research projects than many universities.

The importance of companies' money in research is especially evident when it comes to AI research. There are immense venture capital investments in AI research which made up 57% of the global investments from 2012-2020 (Tricot, 2021). 'The venture capital sector tends to forerun general investment trends, indicating the AI industry is maturing' (Tricot, 2021: 5). While start-up firms based in the USA and China absorbed more than 80% of capital venture investments in 2020, the EU followed in third place with only 4%. Within the EU, Germany and France had the most AI companies accounting for 60% of the EU venture capital investments.

Focusing on universities, Northeastern University (USA) had, with 15 publications, the most on 'Trust in CA' research in the last 15 years, followed by three other institutions with 13 publications, each based in the USA, France, and Germany. In total, three US-based institutions are listed in the top ten list for publications by affiliation. Germany is listed twice, and the other countries are listed just once. There are also three US-based foundations listed in the top ten funding sponsors for trust in CA research publications. Other countries or areas (like China and the EU) are listed twice or less. Also, the author cited the most (Timothy Bickmore) is based in the US. This confirms the huge impact of the USA when it comes to publications on 'Trust in CA'. *There is no indication that the dominance of the US in CA and trust research could be replaced by another country or region based on our research data.*

However, other countries are not to be underestimated as explained in the following. The National Natural Science Foundation in China is the number one funding sponsor for research on 'Trust in CA', followed by the National Science Foundation (USA), and Horizon 2020 (EU). This does not mean that they gave the most money but people funded by these foundations published the majority of all publications in 2009-2024 (Q1) regarding the topic. It also gives an insight into where in the world AI research is a priority for funding sponsors. The list of funding sponsors correlates with the list of publications by country. The USA, China, and EU (especially Germany) dominate the research field so far.

After the USA, Germany seems to have inherited an important role when it comes to 'Trust in CA' research. In Germany, we find the most productive and influential authors. It is not only leading the list with Wolfgang Minker, but it is also the only country with three researchers listed in the top ten. Germany has the second most TP globally and second most TP based on international collaborations (both findings with USA in first place). However, Germany's publication output in 2023/2024 (Q1) decreased from 2021/2022 while all other top-ten countries' publications increased. *Future years will show if Germany can keep its second place in international research when it comes to CA and trust or if other countries will rise up and take its place. Based on our research, candidates could be China, the UK, and France. China has the most TP in 2023/2024 (Q1) and is the number one funding sponsor. The UK almost caught up with Germany in 2023/2024 (Q1) when it comes to TP (however, it is still in fourth place after the USA, China, and Germany) and UK researchers collaborate the most internationally after the USA. France dominates in the top ten affiliation institutions and has two of the top ten researchers.*